\documentclass[%
 reprint,
 amsmath,amssymb,
 aps,
]{revtex4-1}

\usepackage{graphicx}
\usepackage{dcolumn}
\usepackage{bm}
\usepackage{wasysym}

\begin{document}

\preprint{APS/123-QED}

\title{
Enhanced dissociation of H$_2$$^+$ into highly excited states
via laser-induced \\
sequential resonant excitation
}

\author{Kunlong Liu,$^{1}$ Qianguang Li,$^{2}$ Pengfei Lan,$^{1,*}$ and Peixiang Lu$^{1,}$}

\email{Corresponding authors: \\
pengfeilan@hust.edu.cn \\
lupeixiang@hust.edu.cn}

\affiliation{
$^{1}$School of Physics, Huazhong University of Science and Technology, Wuhan 430074, China \\
$^{2}$School of Physics and Electronic-information Engineering, Hubei Engineering University, Xiaogan 432000, China
}%

\date{\today}

\begin{abstract}
We study the dissociation of H$_2$$^+$ in uv laser pulses by solving the non-Born-Oppenheimer time-dependent Schr\"{o}dinger equation as a function of the photon energy $\omega$ of the pulse. Significant enhancements of the dissociation into highly excited electronic states are observed at critical $\omega$. This is found to be attributed to a sequential resonant excitation mechanism where the population is firstly transferred to the first excited state by absorbing one photon and sequentially to higher states by absorbing another one or more photons at the \textit{same} internuclear distance.
We have substantiated the underlying dynamics by separately calculating the nuclear kinetic energy spectra for individual dissociation pathways through different electronic states.
\begin{description}
\item[PACS numbers]
33.80.Rv, 42.50.Hz, 33.80.Wz
\end{description}
\end{abstract}

\pacs{Valid PACS appear here}
\maketitle

\section{Introduction}
Understanding the electronic excitation and nuclear motion in laser-molecular interaction is of vital importance in controlling the formation and fracture of chemical bonds with laser fields \cite{Zewail,Assion,Kling,LanOL,HePRL,ZhuMole}. For more than two decades, many efforts have been made to study the dissociation of small molecules exposed to various laser pulses \cite{Posthumus,MartinS,Karr,Schiller}.
%
Most of the studies are focused on the interplay of the population between the lowest two electronic states of H$_2$$^+$, i.e. the ground ($1s\sigma_g$) and first excited ($2p\sigma_u$) states, because these two states are strongly coupled and isolated from other electronic levels \cite{Moser,Kremer,Liu3}.
For many famous mechanisms, such as bond softening \cite{Bucksbaum}, above-threshold dissociation (ATD) \cite{Giusti}, asymmetric electron localization \cite{Singh}, charge-resonance-enhanced ionization \cite{Zuo,Staudte}, and so on, the H$^+$+H($1s$) dissociation channel of H$_2$$^+$ plays the major role in determining the features of the observed fragmentation phenomena.
Nevertheless, the role of highly excited states has attracted people's attention since
recent works \cite{Gibson,McKenna,Manschwetus,Forre,Erik,Li,Yue,Zhou,Liu} revealed
more details of the molecular fragmentation processes (e.g. multiphoton dissociative excitation \cite{Gibson}, high-order ATD \cite{McKenna}, and Coulomb explosion without double ionization \cite{Manschwetus}) and suggested
that the dissociation through highly excited electronic states might be a ubiquitous phenomenon in laser-molecular interactions.

In an infrared field, however, it is shown that the contribution from highly excited states to the total dissociation is two orders smaller than that from the $1s\sigma_g$ and $2p\sigma_u$ states \cite{Yue}.
This is, on the one hand, due to the weak multiphoton coupling between the $1s\sigma_g$ and highly excited states \cite{Gibson} as well as the ac Stark shifts of the energy levels \cite{Zhu}. On the other hand,
the high ionization rate of highly excited states would also lead to the decrease of the population of those states \cite{Gibson,LiY}.
These facts result that the effects of the population dynamics in high-lying dissociative states would be largely weakened and drowned by the effect of the H$^+$+H($1s$) channel, making the study or control of the high-lying dissociative population dynamics difficult.

In this paper, we study the dissociation of H$_2$$^+$ in uv pulses and report a novel fragmentation process which leads to significantly enhancement of the dissociative population in highly excited states.
In contrast to the previous study \cite{He} where the population is transferred to higher electronic states via single-photon excitation,
the high-lying population in the present study is created through a sequential resonant excitation (SRE) mechanism: For a critical photon energy, the population would be first transferred to the $2p\sigma_u$ state and sequentially to higher states at the \textit{same} internuclear distance.
Our results show that, by taking advantage of the strong coupling between the $1s\sigma_g$ and $2p\sigma_u$ states,
the multiphoton transition through the SRE process
exhibits much higher excitation rate than
the direct multiphoton transition.
The underlying mechanism has been verified by the nuclear kinetic energy release (KER) spectra for individual dissociative electronic states.

\begin{figure*}
\centering\includegraphics[width=17.5cm]{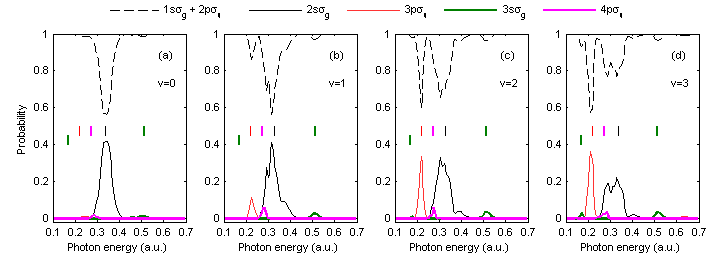}
\caption{\label{fig1}
The probabilities of the dissociation into highly excited electronic states (solid curves) and the total probability of the lowest two states (dashed curves) as a function of photon energy of the pulse for the interaction of H$_2$$^+$ ($v=0$--3) with the uv pulses. The pulse intensity is $3\times10^{13}$ W/cm$^2$ and the pulse duration is 25 optical cycle. The fixed vertical short lines indicate the approximate locations of the dissociation enhancements.
}
\end{figure*}

\section{Theoretical model}
For numerical simulations we have
solved the non-Born-Oppenheimer time-dependent
Schr\"{o}dinger equation (TDSE) for a reduced-dimensionality model of H$_2$$^+$ \cite{H2j,soft}.
The model consists of one-dimensional motion of the nuclei and one-dimensional motion of the electron, and the electronic and nuclear motions are restricted along the polarization direction of the linearly polarized laser pulse.
To date, this model has been widely used to study and identify the molecular fragmentation in strong fields \cite{Liu,Liu2,mioni,Picon,Yue,Madsen,Silva}.
Within this model, the length gauge TDSE can be written as (atomic units are used throughout unless otherwise indicated)
$
i\frac{\partial}{\partial t}\Psi(R,z;t)=[H_0+\varepsilon(t)z]\Psi(R,z;t),
$
where
$
H_0=-\frac{1}{m_p}\frac{\partial ^2}{\partial R^2}
-\frac{1}{2}\frac{\partial ^2}{\partial z^2}
+\frac{1}{R}+V_{e}(z,R)
%
$
%
%
with $V_e(z,R)$ being the improved soft-core potential that reproduces the exact $1s\sigma_g$ potential curve in full dimensions \cite{soft}.
Here, $R$ is the internuclear distance, $z$ is the electron position measured from the center-of-mass of the protons, and $m_p$ is the mass of the proton.
The laser electric field is given by $\varepsilon(t)=\varepsilon_0\sin^2(\pi t/T_d)\sin(\omega t)$ ($0<t<T_d$) with $T_d$, $\omega$, and $\varepsilon_0$ being the full pulse duration, the central frequency, and the peak electric field amplitude, respectively.

The TDSE is solved on a grid by using the Crank-Nicolson split-operator method with a time step of $\Delta t=0.04$ a.u.. The grid ranges from 0 to 40 a.u. for $R$ and from $-200$ to 200 a.u. for $z$, with grid spacings of $\Delta R=0.05$ a.u. and $\Delta z=0.2$ a.u..
After the pulse is off, the wave function of the $j$th electronic state can be obtained by projecting the final wave function to the electronic bound state $\Psi_e^j(z;R)$ at each fixed internuclear distance, i.e.
\begin{eqnarray}
\Psi_D^j(z;R)=\langle\Psi(z;R)|\Psi_e^j(z;R)\rangle\Psi_e^j(z;R),
\end{eqnarray}
where $j$ equals to $0,1,2,...$ and indicates the ground state, the first excited state, the second excited state, and so on.
Then the integration of $\Psi_D^j(R,z)$ produces
the probability of the population in the $j$th electronic state.

To verify the dynamical mechanism of the dissociation, the nuclear KER spectrum for the dissociation channel through a specific electronic state is needed. In the present work, we calculate the channel-specific KER spectra on the basis of the resolvent technique \cite{schafer}. In detail, a channel-specific energy window operator is defined by
\begin{eqnarray}
\hat{W}^j(E_N)=\delta^{2k}/[(\hat{H}^j_N-E_N)^{2k}+\delta^{2k}],
\end{eqnarray}
where
$
\hat{H}_N^j=-\frac{1}{m_p}\frac{\partial ^2}{\partial R^2}+[V_N^j(R)-V_N^j(R=\infty)]
$
with $V_N^j(R)$ being the $R$-dependent potential energy of the $j$th electronic state of H$_2$$^+$.
Then,
the probability distribution for the nuclei having a kinetic energy $E_N$ and the electron being in the $j$th state is extracted from $\Psi_D^j(R,z)$ by applying the energy window operator at each $z$, i.e.
\begin{eqnarray}
p^j(E_N;z)=
\langle\Psi_D^j(R;z)|\hat{W}^j(E_N)|\Psi_D^j(R;z)\rangle.
\end{eqnarray}
Finally, the probability density at $E_N$ of the KER spectrum is given by
\begin{eqnarray}
\rho^j(E_N)=\frac{1}{c}\int p^j(E_N,z)dz
\end{eqnarray}
with $c=\delta\frac{\pi}{k}\csc(\frac{\pi}{2k})$ \cite{Catoire}. In our simulation, we use the parameters $\delta=0.004$ and $k=2$.



%
\section{Results and discussion}
Figure~\ref{fig1} shows
the probabilities of the dissociation into highly excited electronic states (solid curves) and the total probability of the lowest two states (dashed curves)
as a function of pulse frequency $\omega$ for the interaction of H$_2$$^+$ with the uv pulse. The first four vibrational states ($v=0$--3) of H$_2$$^+$ ($1s\sigma_g$) are chosen as the initial states of Figs. 1(a)--1(d), respectively. In the present simulation, the pulse intensity of $3\times10^{13}$ W/cm$^2$ and $T_d=25(2\pi/\omega)$ are used and
$\omega$ ranges from 0.114 to 0.7 a.u. (corresponding to the pulse wavelengths from 400 to 65 nm). Under the pulse parameters that we use here, the ionization is found to be negligible; thus, we will focus on the population of the electronic bound states.


\begin{figure}
\centering\includegraphics[width=8.5cm]{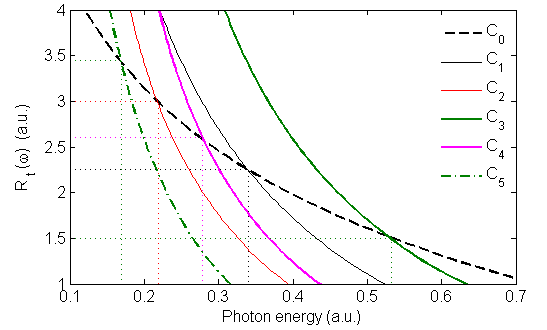}
\caption{\label{fig2}
The transition location $R_t(\omega)$ as a function of photon energy $\omega$ for the resonant transition channels $C_0$--$C_5$ given in the text. The dotted lines illustrate the coordinates of the crossings.
}
\end{figure}

As shown in Fig. 1, the total probability of the $1s\sigma_g$ and $2p\sigma_u$ states (dashed curves) is not conserved as the photon energy varies, indicating that the interplay including only the lowest two states would no longer account for the underlying dynamics even in the case of negligible ionization. More surprisingly,
the remarkable enhancements of the dissociation into the highly excited states ($2s\sigma_g$, $3p\sigma_u$, $3s\sigma_g$, and $4p\sigma_u$) are observed at some critical photon energies.
Moreover, the position of the enhancement for each high-lying state is found to be almost independent on the initial states, whereas the maxima of the dissociation probabilities of different excited states exhibit different tendencies as the vibrational state rises. For instance, the maximum probability of $2s\sigma_g$ (thin black curve) tends to decrease with the vibrational states, but the tendency for $3p\sigma_u$ (thin red curve) is opposite.

In order to understand the features of the anomalous dissociation probability shown in Fig. 1, here we define the transition location $R_t(\omega)$, which represents the internuclear distance where the resonant excitation channel from $1s\sigma_g$ to the excited state opens, as a function of the photon energy $\omega$. Then, based on the $R$-dependent potential energies of the electronic states of H$_2$$^+$, we calculated the transition location $R_t(\omega)$ for the following transition channels [due to parity considerations, coupling between $1s\sigma_g$ and excited gerade (ungerade) electronic states only occurs in even (odd) numbers of photons]:
\begin{eqnarray}
&C_0:\ & 1s\sigma_g+\omega\rightarrow 2p\sigma_u ,\nonumber \\
&C_1:\ & 1s\sigma_g+2\omega\rightarrow 2s\sigma_g ,\nonumber \\
&C_2:\ & 1s\sigma_g+3\omega\rightarrow 3p\sigma_u ,\nonumber \\
&C_3:\ & 1s\sigma_g+2\omega\rightarrow 3s\sigma_g ,\nonumber \\
&C_4:\ & 1s\sigma_g+3\omega\rightarrow 4p\sigma_u ,\nonumber \\
&C_5:\ & 1s\sigma_g+4\omega\rightarrow 3s\sigma_g .\nonumber
\end{eqnarray}
The results have been shown in Fig.~\ref{fig2}. One can see that the curve of $C_0$ (dashed) and other curves intersect at different coordinates, respectively, as indicated by the dotted lines.
Note that the crossing of two curves of $R_t(\omega)$ in Fig. 2 means that, under the photon energy of the crossing, the two represented transition channels would open at the same internuclear distance.
By comparing the results of Figs. 1 and 2, we find that the abscissa values of the crossings in Fig. 2 are approximately equal to those critical photon energies that lead to the enhanced dissociation shown in Fig. 1.
%
Thus, we suggest that, if the resonant excitation channels to the $2p\sigma_u$ state and to the other highly excited state occur at the same internuclear distance and under the same photon energy, the dissociative population of the corresponding state would be significantly increased.

\begin{figure}
\centering\includegraphics[width=8.5cm]{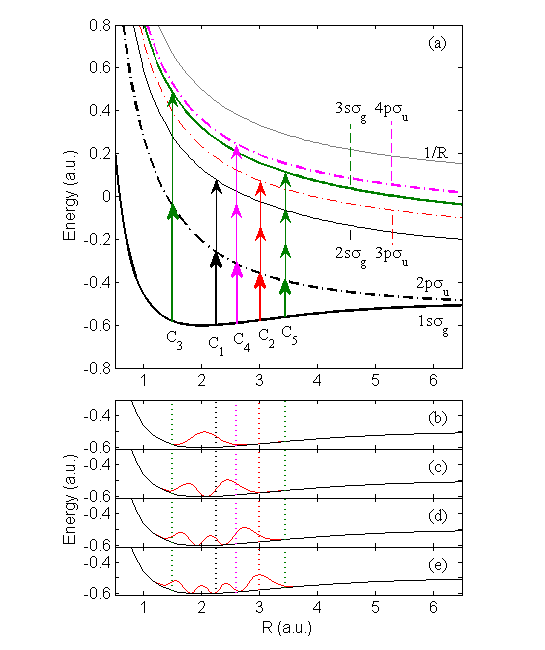}
\caption{\label{fig3}
(a) Illustration of the sequential resonant excitation mechanism.  (b)--(e) The nuclear wave packet profiles on the $1s\sigma_g$ curve for the first four vibrational states of H$_2$$^+$.
}
\end{figure}

\begin{figure*}
\centering\includegraphics[width=15cm]{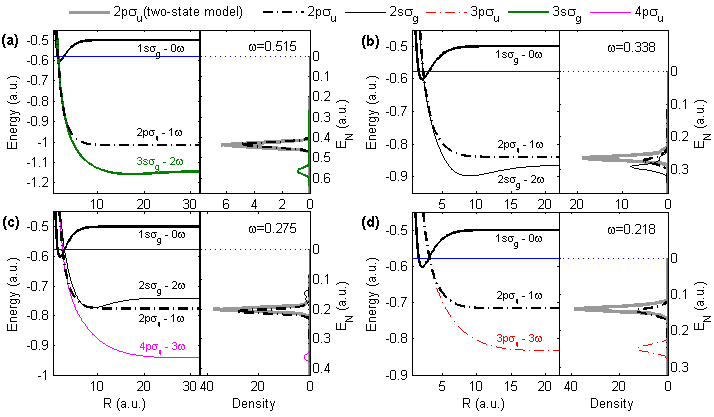}
\caption{\label{fig4}
The dressed molecular potential curves (left panels) and the channel-specific KER spectra of the dissociation into the first five excited electronic states (right panels) in uv pulses. The pulse frequencies are given in the corresponding panels and the other pulse parameters are the same as in Fig. 1. The solid horizontal lines in the left panels indicate the initial vibrational energy of H$_2$$^+$ ($v=2$).
}
\end{figure*}

Based on the above analysis, we now reveal the underlying physical mechanism with the diagram of the molecular potential curves.
Figure~\ref{fig3}(a) illustrates the resonant transition channels $C_1$--$C_5$ that lead to the enhanced dissociation into the highly excited states. The critical photon energies and transition locations for these channels are given by the respective crossings shown in Fig. 2.
It can be seen in Fig. 3(a) that the first photon absorbtion of each channel overlaps with the corresponding $C_0$ channel. In this situation, the direct multiphoton transition
becomes a two-step transition process; that is, the molecule is firstly excited to $2p\sigma_u$ by absorbing one photon and sequentially to higher states by absorbing another one or more photons at the same internuclear distance. We call this process the \textit{sequential resonant excitation} (SRE).
For the first step of the excitation,
due to the strong coupling between the lowest two states, remarkable population would be transferred from $1s\sigma_g$ to $2p\sigma_u$.
Then, for the second step, because the coupling of $2p\sigma_u$ to the higher states is stronger than that of the $1s\sigma_g$ state,
the excited population in $2p\sigma_u$ would be efficiently transferred to the higher state before it dissociates to larger internuclear distance. As a result, the multiphoton transition through the SRE process
would exhibit much higher excitation rate than
the direct multiphoton transition.

Furthermore, besides the excitation rate, the distribution of the initial nuclear wave packet would affect the yield of the excited population \cite{Liu}.
In Figs. 3(b)--3(e), we illustrate the nuclear wave packet profiles on the $1s\sigma_g$ curve for the first four vibrational states of H$_2$$^+$.
For $v=0$, the wave packet concentrates around $R=2$ a.u., so only the dissociative population through the $C_1$ channel is significantly enhanced [see Fig. 1(a)].
For higher vibrational state, the wave packet distribution expands to a wider distribution in $R$ dimension.
Thus the enhancements of other SRE channels gradually become pronounced. In contrast, as the vibrational state rises, the wave packet distribution around the $C_1$ channel is decreased and becomes modulated; thus, the maximum dissociation probability of $2s\sigma_g$ decreases and the modulation structure appears in the
probability curve (thin black) of $2s\sigma_g$ [see Fig. 1(d)].

Next, in order to substantiate the SRE mechanism proposed above, we calculated the channel-specific KER spectra of the dissociation into the first five excited electronic states ($2p\sigma_u$, $2s\sigma_g$, $3p\sigma_u$, $3s\sigma_g$, and $4p\sigma_u$) by using the resolvent technique.
The results for the interaction of H$_2$$^+$ ($v=2$) with the uv pulses of four critical photon energies have been shown in the right panels of Figs.~\ref{fig4}(a)--4(d), including the KER spectra of the $2p\sigma_u$ state (thick gray curves) obtained from the two-state model.
The chosen photon energies, i.e. $\omega=0.515$, 0.338, 0.275, and 0.218 a.u., would trigger the SRE in the $C_3$, $C_1$, $C_4$, and $C_2$ channels, respectively, as shown in Fig. 3(a).
For comparison, the molecular potential curves dressed by the corresponding photon energies are depicted in the left panels of Figs. 4(a)--(d). Note that only the responsible dressed potentials are plotted.

As shown in Fig. 4, if only the lowest two states are considered, the peak positions of the KER spectra of $2p\sigma_u$ (thick gray curves) are in good agreement with the predictions of the $2p\sigma_u-1\omega$ curves.
However, as long as the higher excited states are taken into account, the KER spectra of $2p\sigma_u$ (thick dash-dotted curves) deviate from those of the two-state model.
Compared to the two-state model,
the dissociation yields of $2p\sigma_u$ are lower and the peak positions are shifted in the TDSE calculation.
Such deviations indicate that the dissociative population, which was supposed to dissociate along the $2p\sigma_u$ curve after the single-photon transition, have been partially transferred to higher states before it begins to dissociate. As a result,
in addition to the spectra of $2p\sigma_u$, the KER spectra peaks of other excited states can also be observed in Fig. 4. Moreover, the KER spectra of the highly excited states are found to be in good agreement with the predictions of the responsible dressed potential curves. These results demonstrate that the enhanced dissociation into highly excited states arises from the SRE process.

Additionally, we notice that there is an enhancement in the KER spectrum for the $2s\sigma_g$ state (thin black curve) in Fig. 4(c). According to the dressed potential curves in the left panel, the $2s\sigma_g-2\omega$ and $2p\sigma_u-1\omega$ curves overlap around $R=8.5$ a.u.. When the dissociating population on $2p\sigma_u$ passes the overlapped region, it would be partially excited to $2s\sigma_g$ via absorbing one photon, resulting in the enhancement of the dissociation yield of $2s\sigma_g$. However, due to the delay of the second excitation step, fewer electric field is left to trigger the coupling; thus, the enhancement for $2s\sigma_g$ is weaker than that in Fig. 4(b).

%


\section{Conclusion}
In conclusion, we have studied the dissociation of H$_2$$^+$ in uv pulses as a function of the photon energy of the pulse.
Our results show that
the dissociation into highly excited electronic states
provides a significant contribution at some critical photon energies.
This anomalous phenomenon is attributed to the SRE process in which
considerable population is firstly transferred to the first excited state and sequentially to higher states at the same internuclear distance.
The underlying mechanism has been verified by
the nuclear KER spectra of the dissociation pathways through different electronic states.
Though our present study focuses on the simplified model of H$_2$$^+$, the essential dynamics of SRE should generalized to more complicated molecular systems. In future studies, the effect of the high-lying population dynamics on the molecular fragmentation could be amplified via the SRE process. Moreover, the SRE process would open a feasible access to achieve efficient control of electron localization in highly excited states.

%

\section*{Acknowledgment}
This work was supported by the National Natural Science Foundation of China under Grants No. 11234004 and No.
61275126, the 973 Program of China under Grant
No. 2011CB808103, and the China Postdoctoral Science Foundation under Grant No. 2014M552028. Numerical simulations presented in this paper were partially carried out using the High Performance Computing Center experimental testbed in SCTS/CGCL (see http://grid.hust.edu.cn/hpcc).

\end{document}